\documentclass[%
aip,
jcp,%
amsmath,amssymb,
reprint,%
longbibliography,
floatfix,
nofootinbib]{revtex4-2}
\usepackage{graphicx} 
\usepackage{hyperref}
\hypersetup{colorlinks=true,
	linkcolor=magenta,
	allcolors=magenta}
\usepackage[utf8]{inputenc}
\usepackage[outercaption]{sidecap} 
\usepackage{sidecap} 
\sidecaptionvpos{figure}{c}
\usepackage{upgreek}
\usepackage[version=4]{mhchem}
\usepackage{physics}
\usepackage{siunitx}
\DeclareUnicodeCharacter{2212}{\ensuremath{-}}

\renewcommand{\r}{\mathrm{r}}
\newcommand{\p}{\mathrm{p}}
\newcommand{\A}{\mathrm{A}}
\newcommand{\B}{\mathrm{B}}

\usepackage{float}
\usepackage[caption = false]{subfig}
\AtBeginDocument{\RenewCommandCopy\qty\SI}
\usepackage{xcolor}
\newcommand{\highlight}[1]{{#1}}

\begin{document}

\setlength{\abovedisplayskip}{4pt}
\setlength{\belowdisplayskip}{4pt}

\title{Extending non-adiabatic rate theory to strong electronic couplings in the Marcus inverted regime}
\author{Thomas P. Fay}
\email{tom.patrick.fay@gmail.com}
\affiliation{Aix Marseille University, CNRS, ICR, 13397 Marseille, France}

\begin{abstract}

Electron transfer reactions play an essential role in many chemical and biological processes. Fermi's Golden rule, which assumes that the coupling between electronic states is small, has formed the foundation of electron transfer rate theory, however in short range electron/energy transfer reactions this coupling can become very large, and therefore Fermi's Golden Rule fails to make even qualitatively accurate rate predictions. In this paper I present a simple modified Golden Rule theory to describe electron transfer in the Marcus inverted regime at arbitrarily large electronic coupling strengths. The theory is based on an optimal global rotation of the diabatic states, which makes it compatible with existing methods for calculating Golden Rule rates that can account for nuclear quantum effects with anharmonic potentials. Furthermore the Optimal Golden Rule (OGR) theory can also combined with analytic theories for non-adiabatic rates, such as Marcus theory and Marcus-Levich-Jortner theory, offering clear physical insight into strong electronic coupling effects in non-adiabatic processes. OGR theory is also tested on a large set of spin-boson models and an anharmonic model against exact quantum dynamics calculations, where it performs well, correctly predicting rate turnover at large coupling strengths. 
Finally, an example application to a BODIPY-Anthracene photosensitizer reveals that strong coupling effects inhibit excited state charge recombination in this system, reducing the rate of this process by a factor of four. Overall OGR theory offers a new approach to calculating electron transfer rates at strong couplings, offering new physical insight into a range of non-adiabatic processes.

\end{abstract}

\maketitle

\section{Introduction}
\vspace{-12pt}
\begin{figure}
    \centering
    \includegraphics[width=0.47\textwidth]{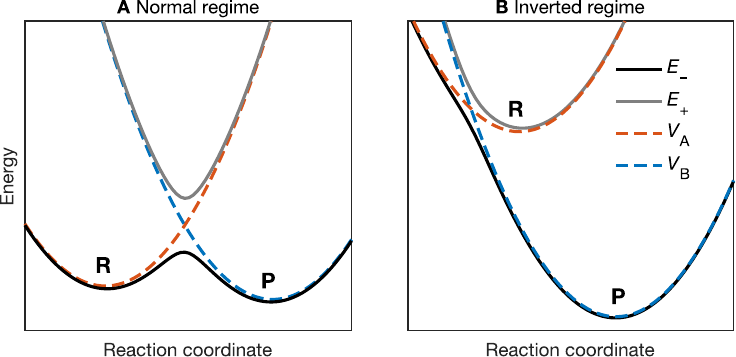}
    \caption{\footnotesize An illustration of the diabatic (dashed lines, $V_\A$ and $V_\B$) and adiabatic/Born-Oppenheimer (solid lines, $E_+$ and $E_-$) potential energy curves for an electron transfer reaction A) in the Marcus normal regime and B) in the Marcus inverted regime. The reactant and product wells are also labelled with R and P respectively.}
    \label{fig-inverted-vs-normal}
\end{figure}

The accurate calculation of electron transfer rates is important in many areas of chemistry, biology and materials science.\cite{blumberger_recent_2015,oberhofer_charge_2017,marcus_electron_1985,santos_models_2022} Most theories for calculating these rate constants assume that the coupling between diabatic electronic states is small, meaning Fermi's Golden Rule perturbation theory can be used.\cite{blumberger_recent_2015} However in many real systems, ranging from photo-active proteins\cite{richter_regulatory_2020} to organic semiconductors\cite{giannini_quantum_2019} to synthetic photocatalysts\cite{wang_insights_2019} the electronic coupling can exceed thermal energy at room temperature,\cite{kubas_electronic_2014} $\sim\!\qty{200}{cm^{-1}}$, sometimes by over a factor of 10.\cite{wang_insights_2019,fay_unraveling_2024} In these cases strong electronic coupling effects can be expected to become important, and the naive application of Fermi's Golden rule can introduce errors of several orders of magnitude in rate constants, particularly for reactions in the Marcus inverted regime.\cite{lawrence_calculation_2019} Thus the aim of this work is to address this challenge and develop a simple theory for the calculation of electron transfer rates at arbitrarily strong electronic coupling in the inverted regime.

For rates in the Golden Rule (GR) limit, there have been impressive developments in recent years, with methods emerging that can accurately predict rate constants including nuclear quantum effects for anharmonic potential energy surfaces.\cite{thapa_nonadiabatic_2019,heller_instanton_2020,heller_semiclassical_2020,lawrence_improved_2020,lawrence_confirming_2020,ananth_path_2022,duke_mean_2016,ranya_multistate_2020,shi_nonradiative_2004}
Many of these methods are obtained using imaginary time path integral or instanton approaches, such as Wolynes theory,\cite{wolynes_imaginary_1987,lawrence_analytic_2018} non-adiabatic instanton methods,\cite{richardson_ring-polymer_2015,heller_instanton_2020} Golden Rule Quantum Transition State Theory (GRQTST),\cite{thapa_nonadiabatic_2019} and the Linear Golden Rule (LGR) method.\cite{lawrence_improved_2020} The success of these methods in the GR limit motivates the development of theories for non-adiabatic rates beyond the weak diabatic coupling regime which are amenable to these path integral and instanton approaches. There have also been significant steps made towards this goal and methods have been developed which can account for strong electronic coupling effects, such as higher-order instanton methods\cite{trenins_nonadiabatic_2022,ansari_instanton_2022,ranya_multistate_2020} and the Non-Adiabatic Quantum Instanton (NAQI) method,\cite{lawrence_general_2020} as well as the development of an interpolation formula technique.\cite{lawrence_calculation_2019}

Existing methods for calculating strong-coupling effects on electron transfer rates generally use the fact that in the Marcus normal regime, where decreasing the free energy change of reaction lowers the activation energy, the reaction can be described as adiabatic, with the electron transfer occurring continuously on a single electronic state (this is illustrated in Fig.~\ref{fig-inverted-vs-normal}A).\cite{lawrence_calculation_2019} This means strong-coupling effects can be accounted for in the normal regime by using established techniques for calculating rate constants on single electronic surfaces,\cite{lawrence_path_2020} such as ring polymer molecular dynamics (RPMD),\cite{craig_chemical_2005} centroid molecular dynamics (CMD),\cite{geva_quantum-mechanical_2001} and instanton methods.\cite{miller_semiclassical_1975,richardson_derivation_2016} However the Marcus inverted regime, where decreasing the reaction free energy increases the activation barrier, still poses a challenge, and the aforementioned methods cannot be applied in this regime (with the possible exception of the higher-order instanton approach\cite{trenins_nonadiabatic_2022}). This is because in the inverted regime, even when diabatic coupling is strong, the reactant and product states still correspond to different adiabatic electronic states, so in this regime the reaction always involves non-adiabatic transitions between different electronic states even in the strong electronic coupling limit.\cite{lawrence_calculation_2019} This is illustrated in Fig.~\ref{fig-inverted-vs-normal}. Direct semiclassical non-adiabatic dynamics approaches, such as surface hopping\cite{mannouch_mapping_2023,lawrence_recovering_2023,runeson_multi-state_2023,runeson_exciton_2024,subotnik_understanding_2016,jain_surface_2015} and mean-field-type methods,\cite{kelly_accurate_2015,pfalzgraff_nonadiabatic_2015,amati_quasiclassical_2022,mulvihill_road_2021} could provide access to strong coupling effects deep in the inverted regime. But these approaches rely on a classical approximation for the motion of nuclei, and in the inverted regime nuclear quantum effects can enhance rates by many orders of magnitude.\cite{lawrence_calculation_2019,lawrence_improved_2020,fay_unraveling_2024} Furthermore rates of processes in the inverted regime can be very slow, with lifetimes exceeding \qty{1}{\mu s} being typical. These factors pose a significant limitation on the application of direct semiclassical dynamics methods to calculate rates in the inverted regime for real molecular systems. Exact quantum dynamical methods can be used to calculate rates for certain simplified Hamiltonians, such as the (Multi-Layer) Multi-Configurational Time-Dependent Hartree method,\cite{wang_multilayer_2003,wang_calculation_2006} Hierarchical Equations of Motion,\cite{tanimura_time_1989,tanimura_numerically_2020} and real-time path-integral methods,\cite{makri_blip_2014,topaler_path_1996} but these methods cannot be applied in general anharmonic systems.

Given that in real systems electronic couplings can be very large for short-range electron transfer,\cite{fay_unraveling_2024,wang_insights_2019,ziogos_ultrafast_2021} there is clearly a need to address strong diabatic coupling effects in the inverted regime. \highlight{Currently the only existing method that can potentially be applied to these problems is Trenins and Richardson's fourth-order instanton method (which was originally developed for the normal regime),\cite{trenins_nonadiabatic_2022} although the reformulation of their theory for the inverted regime has not yet been presented.} Ideally any new theory for this regime should synergise with existing imaginary time path integral and instanton approaches, to enable the inclusion of nuclear quantum effects, which are known to be very important in the inverted regime. In this paper, I outline a theory, based on an optimal rotation of diabatic states, which enables the calculation of electron transfer rates deep in the inverted regime with arbitrarily large diabatic coupling, and which is directly compatible with existing methods for calculating non-adiabatic rates in the Golden Rule regime. I also show how this approach can be used to obtain analytic modified Marcus theory and Marcus-Levich-Jortner theory rate expressions including strong-coupling effects in the inverted regime, and demonstrate how in a BODIPY-Anthracene photosensitizer strong-coupling effects inhibit charge recombination.
\vspace{-25pt}
\section{Theory}
\vspace{-15pt}
\subsection{Non-adiabatic rates}
\vspace{-10pt}
The aim of this work is to find a rate expression for the transfer between diabatic electronic states $\A$ and $\B$. The Hamiltonian for this system is given by
\begin{align}
    \hat{H} = \dyad{\A}\hat{H}_\A + \dyad{\B} \hat{H}_\B + \hat{\Delta} (\dyad{\A}{\B} + \dyad{\B}{\A})
\end{align}
where $\hat{H}_J = \hat{T} + V_J(\hat{\vb*{q}})$ is the diabatic state Hamiltonian for state $J = \A,\B$ and $\hat{\Delta} = \Delta(\hat{\vb*{q}})$ is the diabatic coupling between these states. The diabatic potentials $V_J({\vb*{q}})$ and the coupling ${\Delta}({\vb*{q}})$ are both assumed to be real-valued functions of the nuclear positions $\vb*{q}$. We will be interested in the limit where $\Delta(\vb*{q})$ can be arbitrarily large.

In order to find the rate constant for the transition from $\A$ to $\B$, we first define the reactant and product side operators, which measure the populations of the reactants and products, as $\hat{P}_{\r} = \dyad{\A}$ and $\hat{P}_{\p} = \dyad{\B}$ respectively. These operators define a diabatic state dividing surface between the reactant and product. In the long time limit the expectation values of observables should obey first-order kinetic equations
\begin{subequations}
\begin{align}
    \dv{t}\ev{P_\r(t)} &= -k\ev{P_\r(t)} + k' \ev{P_\p(t)} \\
    \dv{t}\ev{P_\p(t)} &= +k\ev{P_\r(t)} - k' \ev{P_\p(t)}.
\end{align}
\end{subequations}
With this the forward rate constant can be found as\cite{lawrence_calculation_2019}
\begin{align}
    k = \lim_{t\to\infty} \frac{\dv{t}\ev{P_{\p}(t)}}{1-\ev{P_{\p}(t)}/\ev{P_{\p}(\infty)}}.
\end{align}
$\ev*{P_{\p}(t)} = \Tr[e^{+i \hat{H}t/\hbar}\hat{P}_\p e^{-i \hat{H}t/\hbar} \hat{P}_\r^{\mathrm{K}}]/Q_\r$ is the Kubo-transformed side-side correlation function, where the Kubo-transformed operator $\hat{O}^{\mathrm{K}}$ is given by $\hat{O}^{\mathrm{K}} = (1/\beta)\int_0^\beta e^{-(\beta-\lambda)\hat{H}} \hat{O} e^{-\lambda \hat{H}}$ and the reactant partition function is $Q_\r = {\Tr}[\hat{P}_\r e^{-\beta \hat{H}}]$. If the side-side correlation function derivative plateaus on a time $t_\p$ and there is minimal population transfer up to this time,then the rate constant can be simplified to\cite{chandler_introduction_2009}
\begin{align}
    k \simeq \frac{1}{Q_\r} \lim_{t\to t_\p} \dv{t}\Tr[e^{+i \hat{H}t/\hbar}\hat{P}_\p e^{-i \hat{H}t/\hbar} \hat{P}_\r^{\mathrm{K}}].
\end{align} 
With some simple manipulations, the rate constant can alternatively be written in terms of the flux-flux correlation function as
\begin{align}\label{eq-ff-rate}
    k \simeq \frac{1}{2 Q_\r} \int_{-t_\p}^{t_{\p}} \Tr[e^{+i \hat{H}t/\hbar}\hat{F}e^{-i \hat{H}t/\hbar} \hat{F}^{\mathrm{K}}]\dd{t},
\end{align}
where the flux operator $\hat{F} = \frac{i}{\hbar} [\hat{H},\hat{P}_{\r}]$. The standard Golden rule rate expression can be obtained by expanding this to lowest order in the coupling term $\hat{\Delta}$. This can be achieved by partitioning the Hamiltonian into $\hat{H}_0 + \hat{V}$ with $\hat{H}_0 = \dyad{\A}\hat{H}_\A + \dyad{\B} \hat{H}_\B$ and $\hat{V} = \hat{\Delta} (\dyad{\A}{\B} + \dyad{\B}{\A})$ and noting that $\hat{F} = (i/\hbar)\hat{\Delta}(\dyad{\B}{\A} - \dyad{\A}{\B}) $ and $e^{z \hat{H}} = e^{z\hat{H}_0} + \mathcal{O}(\Delta)$, which yields the Golden Rule rate expression,\cite{richardson_non-oscillatory_2014}
\begin{align}\label{eq-kGR}
    k_\mathrm{GR} \!\!=\!\! \frac{1}{\hbar^2 Q_{\A}} \int_{-\infty}^{\infty} \!\!\!\Tr_\mathrm{n}[e^{-\beta \hat{H}_\A -i \hat{H}_\A t/\hbar}\hat{\Delta}e^{+i \hat{H}_\B t/\hbar}\hat{\Delta}]\dd{t}\!,
\end{align}
where $\Tr_\mathrm{n}[\cdots]$ denotes a trace just over nuclear degrees of freedom, and $Q_{J} = \Tr_\mathrm{n}[e^{-\beta \hat{H}_J}]$ for $J = \A,\B$. 
The Golden Rule rate is accurate in the limit where the diabatic coupling $\Delta$ is small, but can deviate significantly from the exact rate beyond this limit. The aim of the following is to find an alternative approximation to the full rate constant expression, Eq.~\eqref{eq-ff-rate}, which accurately captures strong diabatic coupling effects.
\vspace{-30pt}
\subsection{A modified Golden Rule rate}
\vspace{-10pt}
For any sensible choice of dividing surface that does not radically change the character of the reactant and product states, the exact rate constant $k$ is unchanged. This is because the rate constant is simply the slowest relaxation rate of the entire system. Thus it is possible to choose an alternative dividing surface, which defines $\hat{P}_{\r}$ and $\hat{P}_{\p} = 1 - \hat{P}_\r$, for which the coupling between states is minimised, and perturbation theory can therefore be applied more accurately. Here I will define the side operators using rotated electronic states $\ket{\psi_\pm(\theta(\vb*{q}))}$ (although the explicit $\vb*{q}$ dependence of these states will be dropped shortly), such that the side operators are given by,
\begin{align}
    \hat{P}_{\r} &= \dyad{\psi_{+}(\theta({\vb*{q}}))}
\end{align}
\begin{align}
    \hat{P}_\p &= \dyad{\psi_-(\theta(\vb*{q}))}
\end{align}
where the rotated states are defined as
\begin{align}
    \ket{\psi_+(\theta(\vb*{q}))} &= \ket{\A}\cos(\theta(\vb*{q})) + \ket{\B}\sin(\theta(\vb*{q}))\\
    \ket{\psi_-(\theta(\vb*{q}))} &= -\ket{\A}\sin(\theta(\vb*{q})) + \ket{\B}\cos(\theta(\vb*{q})).
\end{align}
If the rotation angle $\theta(\vb*{q})$ is chosen to diagonalise the Born-Oppenheimer Hamiltonian $\hat{H} - \hat{T}$ then these states are the adiabatic states. In this case the rotation angle is given by
\begin{align}
    \tan(2\theta(\vb*{q})) = \frac{2\Delta(\vb*{q})}{V_\A(\vb*{q}) - V_\B(\vb*{q})}
\end{align}
and the eigenvalues of $\hat{H} - \hat{T}$ are given by
\begin{align}
    E_{\pm}(\vb*{q}) = \frac{V_\A(\vb*{q})+V_\B(\vb*{q})}{2} \pm \frac{\sqrt{4\Delta(\vb*{q})^2 + (V_\A(\vb*{q}) - V_\B(\vb*{q}))^2}}{2}
\end{align}
but in this basis the kinetic energy operator $\hat{T}$ acquires off-diagonal elements which become very large in regions of position space where the adiabatic energies are close. This suggests that choosing the adiabatic state dividing surface may not be the optimal choice. 

The large off-diagonal elements of the kinetic energy operator in the adiabatic basis originate in the position dependence of the electronic states. One simple way to remove these is to choose a global rotation angle, $\theta(\vb*{q}) \to \theta$, to define the rotated diabatic basis states $\ket{\pm} \equiv \ket{\psi_{\pm}(\theta)}$ which can then be used to define the dividing surface between reactants and products. The Hamiltonian can be written in terms of these globally rotated diabatic states as
\begin{align}
    \hat{H} &= \hat{H}_+ \dyad{+} + \hat{H}_- \dyad{-} + \hat{\Delta}_{+-} (\dyad{+}{-} + \dyad{-}{+})
\end{align}
where the constituent terms are
\begin{align}
    \hat{H}_+ &=  \hat{H}_\A\cos^2\theta + \hat{H}_\B\sin^2\theta  + \hat{\Delta} \sin(2\theta) \label{eq-h+} \\
    \hat{H}_- &=  \hat{H}_\A\sin^2\theta  + \hat{H}_\B\cos^2\theta  - \hat{\Delta} \sin(2\theta) \label{eq-h-}\\
    \hat{\Delta}_{+-} &= \hat{\Delta} \cos(2\theta) + \frac{1}{2}(\hat{H}_\B - \hat{H}_\A)\sin(2\theta).
\end{align}
With these rotated diabatic states, a perturbative expression for the rate constant can be obtained by setting $\hat{H}_0 = \hat{H}_+ \dyad{+} + \hat{H}_- \dyad{-}$ and $\hat{V} = \hat{\Delta}_{+-} (\dyad{+}{-} + \dyad{-}{+})$ and expanding the resulting expression to lowest order in $\Delta_{+-}$, which gives a modified Golden Rule rate expression, $k_\mathrm{GR,\theta}$,
\begin{align}\label{eq-ogr}
    k_\mathrm{GR,\theta} \!=\! \frac{1}{\hbar^2 Q_{+}} \!\int_{-\infty}^{\infty}\! \!\!\!\Tr_\mathrm{n}[e^{-\beta \hat{H}_+ -i \hat{H}_+ t/\hbar}\hat{\Delta}_{+-}e^{+i \hat{H}_- t/\hbar}\hat{\Delta}_{+-}]\!\dd{t}\!,
\end{align}
where $Q_{\pm} = \Tr_\mathrm{n}[e^{-\beta \hat{H}_\pm}]$. 
If $\theta = 0$ this reduces to the standard Golden Rule rate expression, but for $\theta \neq 0$ the coupling term $\hat{\Delta}$ is partially folded into reference Hamiltonian, which in principle enables strong electronic coupling effects to be accounted for in the rate constant. The question that remains is what is the optimal choice of mixing angle $\theta$? This is addressed in the next section.
\vspace{-10pt}
\subsection{The optimal choice of $\theta$}
\vspace{-10pt}
The perturbative expansion of the rate constant in $\Delta_{+-}$ will be most justified when the magnitude of this term is minimised in some sense. There are many potential measures of how strong $\Delta_{+-}$ is, but one simple choice is to minimise the expectation value of the square coupling ${\Delta_{+-}^2}$ in either in the local equilibrium distribution of $\ket{+}$ or $\ket{-}$, $\hat{\rho}_{\pm}^\mathrm{eq} = e^{-\beta \hat{H}_{\pm}}/Q_\pm$. 
\highlight{
Deep in the inverted regime, the fluctuations in $\Delta\hat{V}= \hat{V}_\B-\hat{V}_{\A}$, denoted $\delta \Delta\hat{V} = \Delta\hat{V} - \ev{\Delta\hat{V}}_{\pm}$, will be small compared to $\ev{\Delta V}_\pm^2$, so the $\ev{\delta\Delta\hat{V}^2}_{\pm}$ terms appear in the above can be ignored. Likewise if the fluctuations in the coupling are also ignored (which is not necessary if the Condon approximation is also invoked\cite{lawrence_calculation_2019} and $\hat{\Delta}$ is assumed to be a constant), then $\ev{\Delta_{+-}^2}_{\pm}$ can be approximated as
\begin{align}
    \ev{\Delta_{+-}^2}_{\pm} 
    \approx \left(\ev{\Delta}_\pm \sin(2\theta)+\frac{1}{2}\ev{\Delta V}_\pm \cos(2\theta) \right)^2.
\end{align}
Now assuming $e^{-\beta \hat{H}_+} \approx e^{-\beta \hat{H}_\A}$ and $e^{-\beta \hat{H}_-} \approx e^{-\beta \hat{H}_\B}$ we find that $\ev{\Delta_{+-}^2}_{\pm}$ is given approximately by
\begin{align}
    \ev{\Delta_{+-}^2}_{+} &\approx 
    \left(\ev{\Delta}_\A \sin(2\theta)+\frac{1}{2}\ev{\Delta V}_\A \cos(2\theta) \right)^2 \\
    \ev{\Delta_{+-}^2}_{-} &\approx 
    \left(\ev{\Delta}_\B \sin(2\theta)+\frac{1}{2}\ev{\Delta V}_\B \cos(2\theta) \right)^2,
\end{align}
where $\ev{\cdots }_J = \Tr_\mathrm{n}[\cdots e^{-\beta \hat{H}_J}/Q_J]$. Finally differentiating these expressions with respect to $\theta$ yields values of $\theta$ which minimise the coupling in the A and B equilibrium distributions,
\begin{align}
    \tan(2\theta_J) =  \frac{2\ev{\Delta}_J}{\ev{V_\A - V_\B}_J}
\end{align}
for $J = \A,\B$. Clearly the coupling cannot be minimised simultaneously for the $\ket{+}$ and $\ket{-}$ states with a global choice of $\theta$. This suggests that the optimal value of $\theta$ should lie somewhere in this range.} 
\begin{figure}
    \centering
    \includegraphics[width=0.46\textwidth]{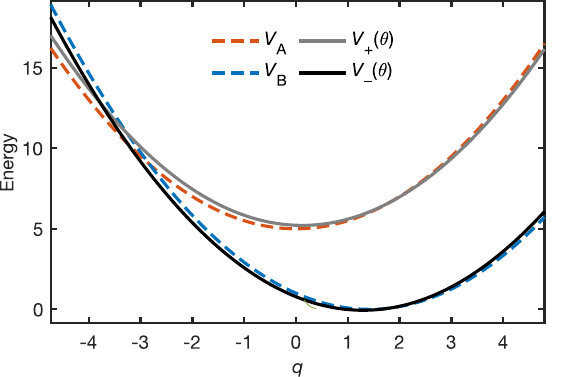}
    \caption{\footnotesize\highlight{An illustration of the rotated diabatic states for a single-mode spin-boson model with $\epsilon =-\Delta A = 5$, $\lambda = 2$, $m\omega^2 = 1$ and $\Delta = 1$ with $\theta = \theta_\A$. } }
    \label{fig-theta-demo}
\end{figure}

In order to obtain an optimal value, we note that the Golden Rule rate is essentially a transition state theory, and the higher order terms in $\Delta_{+-}$ that are neglected correspond primarily to recrossing effects in the Marcus inverted regime,\cite{evans_incorporating_1995,golosov_reference_2001,richardson_non-oscillatory_2014} and therefore $k_{\mathrm{GR},\theta}$ should be an approximate upper bound on the true rate constant in the inverted regime. This motivates defining the optimal value of $\theta$ as the value that minimises the rate in the range $\theta_\B \leq \theta \leq \theta_\A$, so the optimal value of the rate constant is given by
\begin{align}\label{eq-kOGR1}
    k_{\mathrm{GR},\theta^*} = \underset{\theta \in [\theta_\B,\theta_\A]}{\mathrm{min}} k_{\mathrm{GR},\theta}.
\end{align}
Very deep in the inverted regime, assuming $\B$ is the lower energy state, we expect the rate constant to be minimised when either $\theta = \theta_\B \approx 0$ in the limit of small diabatic coupling $\Delta$ or at $\theta = \theta_\A$ in the strong diabatic coupling limit. This motivates choosing the optimal rate constant to be the minimum of $k_{\mathrm{GR},\theta_\A}$ and $k_{\mathrm{GR},\theta_\B}$,
\begin{align}\label{eq-kOGR2}
    k_{\mathrm{GR},\theta^*} \approx \min(k_{\mathrm{GR},\theta_\A},k_{\mathrm{GR},\theta_\B}),
\end{align}
if calculations for the full range of $\theta$ are prohibitive. 

\begin{figure}
    \centering
    \includegraphics[width=0.45\textwidth]{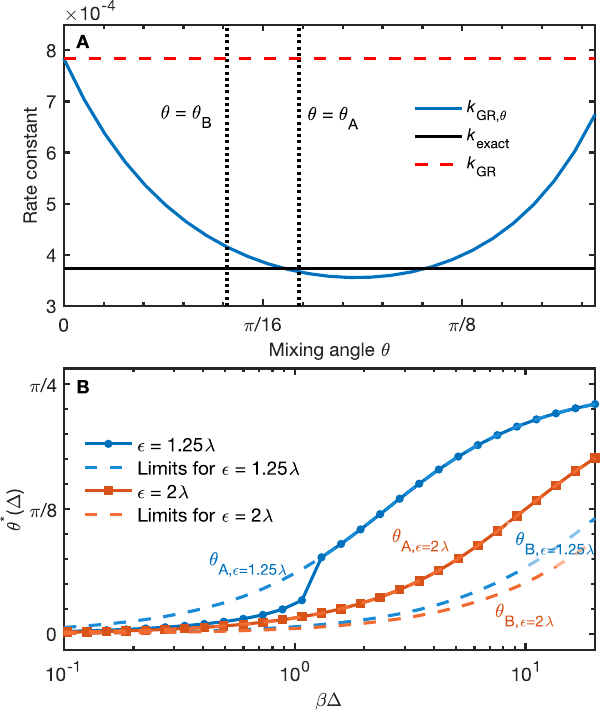}
    \caption{\footnotesize\highlight{A) Behaviour of the $k_{\mathrm{GR},\theta}$ as a function of $\theta$ for a spin-boson model with an underdamped Brownian Oscillator spectral density with $\beta\lambda=20$, $\beta\epsilon = 100$, $\beta\hbar\Omega = 4$, $\beta\hbar\gamma = 4$ and $\beta\Delta = 20$. The exact HEOM rate constant and the GR rate constants are also shown, together with the proposed limits for the optimisation of $k_{\mathrm{GR},\theta}$. B) The optimal values of $\theta$ for an underdamped spin-boson model close to the activation-less case $\epsilon = 1.25\lambda$ and in the inverted regime $\epsilon = 2\lambda$ as a function of $\beta\Delta$. In this example $\beta \hbar\Omega = \beta\hbar\gamma = 4$ and $\beta\lambda = 20$.} }
    \label{fig-theta-opt}
\end{figure}

\highlight{An example of the rotated diabats with $\theta = \theta_\A$ are shown in Fig.~\ref{fig-theta-demo}, where it can be seen that the rotatation raises the energy of the crossing point between diabats and thereby decreases the rate constant, so we see qualitatively how this rotation of the diabats can reduce the rate constant.} \highlight{Fig.~\ref{fig-theta-opt}A shows the behaviour of $k_{\mathrm{GR},\theta}$ as a function of $\theta$ for a spin-boson model (see Sec.~\ref{sec-sb-models} for details) deep in the inverted regime, with $\epsilon = -\Delta A= 5 \lambda$. We see that the rate constant is minimised at $\theta = \theta_\A$ in the range $\theta_\B \leq \theta \leq \theta_\A$, although this is not unconstrained minimum which lies at $\theta > \theta_\A$. This minimum underestimates the exact HEOM rate constant, and the constrained minimum provides a better estimate of the true rate constant than the true minimum. This shows the importance of applying a physically motivated constraint on optimisation of the global rotation angle. Furthermore we see that in the range $\theta_\B \leq \theta \leq \theta_\A$ $k_{\mathrm{GR},\theta}$ is a monotonic function of $\theta$, thus the OGR rate Eq.~\eqref{eq-kOGR1} and the simplified OGR rate Eq.~\eqref{eq-kOGR2} are equivalent, which motivates the simplified form Eq.~\eqref{eq-kOGR2}.}

\highlight{To further explore the behaviour of $\theta^*$, in Fig.~\ref{fig-theta-opt}B the optimal values of $\theta$ are plotted as a function of $\beta\Delta$ for a spin-boson model in the inverted regime (details of the model are given in Sec.~\ref{sec-sb-models}). Deep in the inverted regime, $\epsilon \geq 2 \lambda$, the rate is minimised at $\theta = \theta_B$ (which is true for all deep-inverted regime models explored in this work). Conversely closer to the activation-less case, $\epsilon = 1.25 \lambda$, a transition is observed where the optimal value of $\theta$ changes from close to $\theta_\A$ to $\theta_\B$ as $\Delta$ increases, with $\theta^*$ not lying exactly at one of the bounds for only a relatively small range of values of $\Delta$. This suggests that the simplified optimisation, Eq.~\eqref{eq-kOGR2}, should accurately approximate the full version as long as the problem in question lies far from the Marcus activation-less regime.}

This approximate optimisation condition can be viewed through the lens of the higher-order non-adiabatic rate theory in Ref.~\onlinecite{trenins_nonadiabatic_2022}. Without going into detail, it has been found that within a semiclassical ($\hbar \to 0$) approximation, the dominant term in the fourth-order term in $\Delta$ in the perturbative expansion of the rate constant, denoted $k^{(4)}$, is proportional to $k_{\mathrm{GR}}^2$. This is also true in Zusman theory\cite{zusman_theory_1988} and its generalisations\cite{sparpaglione_dielectric_1988,sparpaglione_dielectric_1988-1}. This means that the square ratio of the fourth order contribution to the GR contribution, $(k^{(4)}/k_{\mathrm{GR}})^2$, is approximately proportional to $k_{\mathrm{GR}}^2$.  
\highlight{Minimising the magnitude of higher-order contributions to the rate should make the lowest order rate constant closer the the true rate. By constraining $\theta$ to lie in the range $\theta_\B \leq \theta \leq \theta_\A$ we ensure that the the approximate minimisation of the fourth-order contribution to the rate constant is constrained to values of $\theta$ which also minimise the coupling between diabatic states and which preserve the character of the reactant and product states.}

The modified coupling term $\hat{\Delta}_{+-}$ contains the diabatic energy gap $\hat{H}_\B - \hat{H}_\A$, which may complicated evaluation of the rate constant. We note however that the energy gap operator can be written as
\begin{align}
    \hat{H}_\B - \hat{H}_\A = (\hat{H}_- - \hat{H}_+) \sec(2\theta) + 2 \Delta \tan(2\theta),
\end{align}
and furthermore any terms proportional to $\hat{H}_- - \hat{H}_+$ do not contribute to the rate constant $k_{\mathrm{GR},\theta}$. \highlight{An intuitation for this at first surprising result can be gained by considering the classical limit. In this limit transitions between rotated diabats only occur at points in configuration space $\vb*{q}$ where diabatic states, $V_{\pm}(\vb*{q})$, are equal in energy, $V_+(\vb*{q}) = V_-(\vb*{q})$, so $ V_-(\vb*{q}) - V_+(\vb*{q})$ contributions to the coupling vanish where diabatic transitions occur and they do not contribute to the rate constant. The generalisation of this to the quantum case is given in Appendix \ref{app-kOGR-simp}.} This means the rate constant can be expressed more simply as
\begin{align}\label{eq-ogr-simp}
    k_\mathrm{GR,\theta} \!=\! \frac{\sec^2(2\theta)}{\hbar^2 Q_{+}} \!\!\int_{-\infty}^{\infty} \!\!\Tr_\mathrm{n}[e^{-\beta \hat{H}_+ -i \hat{H}_+ t/\hbar}\hat{\Delta}e^{+i \hat{H}_- t/\hbar}\hat{\Delta}]\dd{t}\!.
\end{align}
One clear advantage of this modified Golden Rule expression is that it has exactly the same form as the standard Golden Rule rate expression, which means that existing path integral approximations to the GR rate, such as Wolynes theory\cite{wolynes_imaginary_1987,lawrence_analytic_2018}, \highlight{instanton approaches\cite{heller_instanton_2020,trenins_nonadiabatic_2022}} and the linear Golden Rule method,\cite{lawrence_improved_2020} can be directly applied with this method. 
Overall these equations define the Optimal Golden Rule rate (OGR) method. 
\vspace{-10pt}
\subsection{Modified Marcus and Marcus-Levich-Jortner theories}
\vspace{-10pt}
\begin{figure*}[ht]
    \centering
    \includegraphics[width=0.92\textwidth]{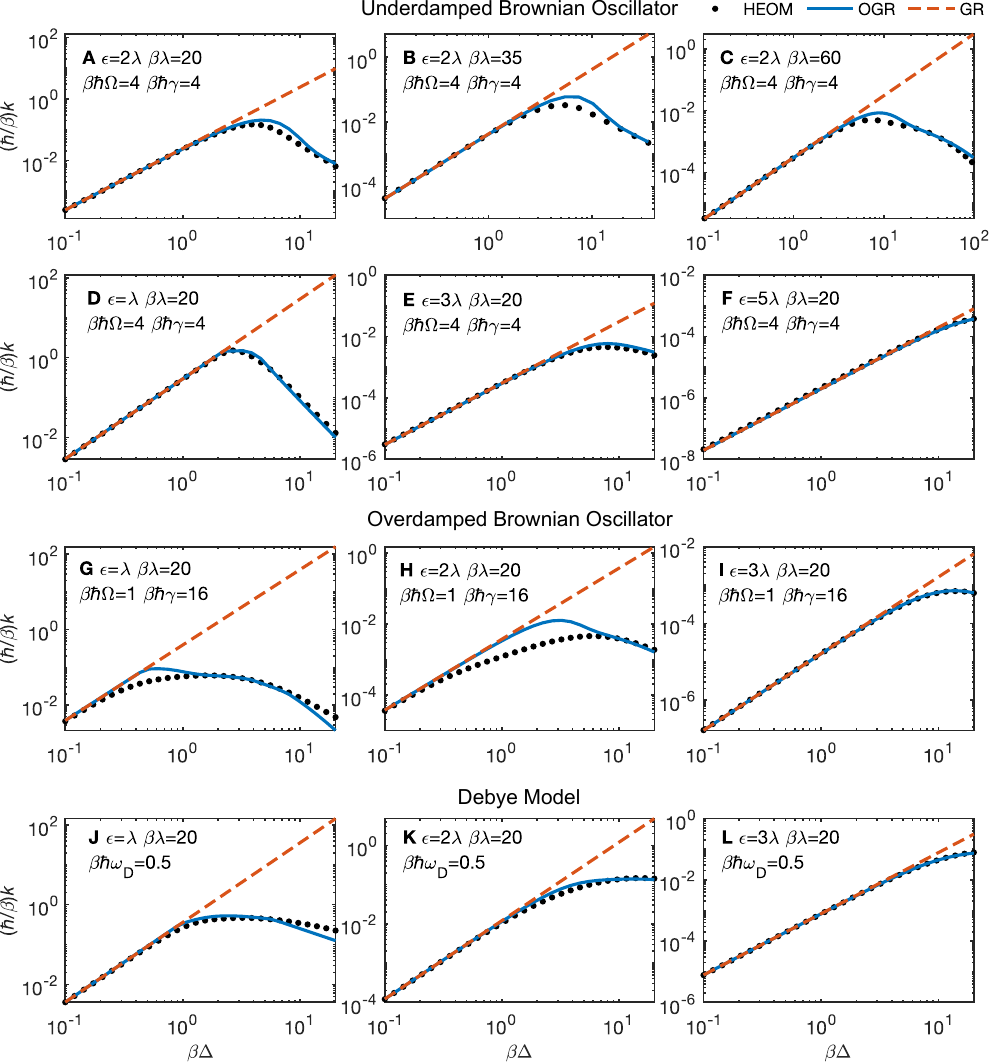}
    \vspace{-10pt}
    \caption{\footnotesize A--I) HEOM, GR and OGR rate constants as a function of $\beta\Delta$ for a range of spin-boson models with a Brownian oscillator spectral density.  J--L) HEOM, FG and OGR rate constants as a function of $\beta\Delta$ for spin-boson models with a Debye spectral density. The model parameters are given in the individual panels. From A--C the reorganisation energy increases with $\epsilon = 2 \lambda$. From D--F, G--I, and J--L the rate moves from the activation-less regime $\epsilon = \lambda$, to deep in the inverted regime, up to $\epsilon = 5 \lambda$ for F and $\epsilon = 3 \lambda$ for I and L. The exact results in C are taken from Ref.~\onlinecite{lawrence_calculation_2019}. }
    \label{fig-sb-rates}
    \vspace{-10pt}
\end{figure*}

It is straightforward to apply the optimal golden rule expression method to spin-boson models, where the diabatic coupling is treated as a constant $\hat{\Delta} = \Delta$ the diabatic Hamiltonians are treated as a set of independent Harmonic oscillators 
\begin{align}\label{eq-spin-boson-pes}
    \hat{V}_{J} = \sum_\alpha \frac{1}{2}m_\alpha \omega_\alpha^2 \left(\hat{q}_\alpha - \frac{c_{J,\alpha}}{m_\alpha \omega_\alpha^2}\right)^2 + \epsilon_J,
\end{align}
Here $m_\alpha$ and $\omega_\alpha$ are the oscillator masses and frequencies, $\epsilon_J$ is the energy of $J = \A,\B$ and $c_{J,\alpha} = \delta_{J,\B}c_\alpha$ quantifies how much each oscillator is displaced by going from diabatic state $\A \to \B$. This is related to the reorganisation energy $\lambda$ for the harmonic oscillators by $\lambda = \sum_\alpha {c_\alpha^2}/{2m_\alpha \omega_\alpha^2}$. The free energy change for the $\A \to \B$ reaction in this model is simply $\Delta A = \epsilon_\B - \epsilon_\A$. \highlight{Note that this definition differs slightly from that often used in the literature, since only state B couples to the bath, but this choice has the advantage of making the definitions of the reorganisation energy in terms of the spectral density (see below) and the diabatic energy differences equivalent.} 

\highlight{Applying the OGR method requires the rotated diabatic Hamiltonians, $\hat{H}_{\pm}$. Given that the rotated diabatic Hamiltonians $\hat{H}_{\pm}$, Eqs.~\eqref{eq-h+} and \eqref{eq-h-}, are linear combinations of the original diabats (with an additional constant shift) it follows that the new oscillator frequencies after rotation are unchanged and only the coefficients coupling $c_\alpha$ are altered. From this it follows that the rotated harmonic potential energy surfaces are also harmonic with the same distribution of frequencies, but with a re-normalised reorganisation energy $\tilde{\lambda}_\theta$ and free energy change $\Delta\tilde{A}_\theta$, which after some simple algebra are found to be}
\begin{align}
    \tilde{\lambda}_\theta &= \lambda \cos^2(2\theta) \\
    \Delta\tilde{A}_\theta &= \Delta{A}\cos(2\theta) - 2 \Delta \sin(2\theta).
\end{align}
Deep in the Marcus inverted regime, the optimised value of $\theta$ is $\theta^* = \theta_\A = (1/2)\atan(2\Delta/|\lambda + \Delta A|)$. The Marcus theory rate\cite{hush_adiabatic_1958,kubo_application_1955,marcus_theory_1956} is obtained from the high-temperature limit of the exact spin-boson rate constant expression. This means that the modified Marcus theory rate is simply given by
\begin{align}\label{eq-mod-marcus}
k_{\mathrm{MT},\theta^*} = \frac{2\pi \Delta^2 \sec^2(2\theta^*)}{\hbar\sqrt{{4\pi\tilde{\lambda}_{\theta^*}k_\mathrm{B}T}}}   \exp(- \frac{(\tilde{\lambda}_{\theta^*} + \Delta\tilde{A}_{\theta^*})^2}{4\tilde{\lambda}_{\theta^*}k_\mathrm{B}T} ).
\end{align}
So overall the modified Marcus expression is simply the standard Marcus theory rate expression with modified coupling, $\Delta\to \Delta\sec(2\theta^*)$, reorganisation energy, $\lambda \to \tilde{\lambda}_{\theta^*}$, and free energy change, $\Delta A \to \Delta \tilde{A}_{\theta^*}$. To lowest order in the coupling $\Delta$, this can be approximated as 
\begin{align}
    k_{\mathrm{MT},\theta^*} \simeq k_{\mathrm{MT},\theta = 0} \exp(-\frac{2 \Delta^2}{|\lambda + \Delta A|k_\B T}), 
\end{align}
where $ k_{\mathrm{MT},\theta = 0}$ is the standard Marcus theory rate constant. This reveals that the first-order effect of strong coupling is to raise the activation barrier, and thus suppress the rate constant.

In Marcus-Levich-Jortner (MLJ) theory the harmonic bath is assumed to consist of a low frequency part, with reorganisation energy $\lambda_\mathrm{o}$, which can be treated with a high temperature approximation and a single high-frequency mode, with frequency $\Omega$ and  reorganisation energy $\lambda_\mathrm{v}$, which is treated with a low temperature approximation. Exactly the same renormalisation of the reorganisation energy and free energy change can be applied in this case, yielding the following modified MLJ theory rate constant,
\begin{widetext}
\begin{align}
    k_{\mathrm{MLJ},\theta^*} \!=\! \frac{2\pi \Delta^2 \sec^2(2\theta^*)}{\hbar\sqrt{{4\pi\tilde{\lambda}_{\mathrm{o},\theta^*}k_\mathrm{B}T}}} \!\sum_{\nu = 0}^{\infty} \!\left(e^{-\tilde{S}_{\theta^*}}\!\frac{\tilde{S}_{\theta^*}^\nu}{\nu!}\!\right) \exp[- \frac{(\tilde{\lambda}_{\mathrm{o},\theta^*} + \Delta\tilde{A}_{\theta^*} + \hbar \Omega \nu )^2}{4\tilde{\lambda}_{\mathrm{o},\theta^*}k_\mathrm{B}T}]
\end{align}
\end{widetext}
where $\tilde{S}_{\theta^*} = \tilde{\lambda}_{\mathrm{v},\theta^*}/\hbar \Omega $ is the renormalised Huang-Rhys factor. Qualitatively one can see that the reduction in the Huang-Rhys factor on renormalisation of the diabats by strong coupling leads to a reduction in tunnelling and therefore a reduction in the rate constant. 

\highlight{Whilst these approximate theories may be qualitatively useful, it is shown in Appendix \ref{app-zus} that the modified Marcus theory is not quantitatively accurate compared to spin-boson models, although it does generally predict the value of $\Delta$ at which turnover occurs fairly accurately. This is due to the neglect of nuclear tunnelling effects in Marcus theory, and as such it is advisable to use the full quantum theory for the spin-boson rate constant with a full spectral density for processes deep in the inverted regime with strong electronic coupling.}
\vspace{-5pt}
\section{Numerical tests}
\vspace{-10pt}
Having presented a theory of strong coupling effects in the Marcus inverted regime in the previous, I now move on to demonstrate the accuracy of this method on a series of model systems, including an anharmonic condensed phase system, for which numerically exact rates can be obtained using the Hierarchical Equations of Motion (HEOM) method. 
\vspace{-10pt}
\subsection{Spin-boson models}\label{sec-sb-models}
\vspace{-10pt}
The first set of model problems considered are spin-boson models, where the model diabatic potentials are given by Eq.~\eqref{eq-spin-boson-pes}. The dynamics of the electronic variables (and therefore also the rate of transfer from A to B), is dictated fully by the spectral density,
\begin{align}
    \mathcal{J}(\omega) = \frac{\pi}{2}\sum_{\alpha} \frac{c_\alpha^2}{m_\alpha \omega_\alpha} \delta(\omega - \omega_\alpha) .
\end{align}
\highlight{The reorganisation energy $\lambda$ is related to the spectral density  by $\lambda = (1/\pi)\int_0^\infty \mathcal{J}(\omega)/\omega\dd{\omega} $.} Both Brownian oscillator and Debye bath models have been considered, with spectral densities given respectively by
\begin{align}
    \mathcal{J}_{\mathrm{BO}}(\omega) &=  \frac{2\lambda\gamma\Omega^2\omega}{(\omega^2 -\Omega^2)^2 + \gamma^2 \omega^2} ,\\
     \mathcal{J}_{\mathrm{D}}(\omega) &=  \frac{2\lambda\omega_\mathrm{D}\omega}{\omega_\mathrm{D}^2 + \omega^2}.
\end{align}
All HEOM calculations (expect those in Fig.~\ref{fig-sb-rates}C, which were taken from Ref.~\onlinecite{lawrence_calculation_2019}) were performed with the \texttt{heom-lab} code using a Matsubara expansion of the bath correlation functions with the truncation scheme outlined in Ref~\onlinecite{lindoy_quantum_2023}. with the low temperature correction scheme outlined in Ref.~\onlinecite{fay_simple_2022}. Further details of the HEOM calculations are given in Appendix~\ref{app-heom}. In all cases the diabatic coupling is treated as a constant $\hat{\Delta} = \Delta$ between a value of \qty{0.1} and around \qty{20} or larger, which corresponds to a range of values typically encountered for molecular systems, between about \qty{20}{cm^{-1}} and \qty{4000}{cm^{-1}} for $T=\qty{298}{K}$. In addition to the spectral density parameters, the reorganisation energy $\lambda$ and the free energy change $\epsilon = \epsilon_\A - \epsilon_\B$ are all varied, to explore a range of models, from the activationless regime, $\epsilon = \lambda$, to deep in the inverted regime $\epsilon = 5 \lambda$.

In Fig.~\ref{fig-sb-rates} the HEOM rate data are plotted together with the GR rate (Eq.~\eqref{eq-kGR}) and the OGR rate (Eq.~\eqref{eq-kOGR1}) as a function of $\beta\Delta$. In general we see that the OGR approximation correctly captures the Golden Rule limit, where $\beta\Delta$ is small, and the subsequent turnover of the rate at larger values of $\beta\Delta$. This is robust to the choice of spectral density, with the OGR approach working for a highly quantum and underdamped case $\beta\hbar\Omega = 4$, $\beta\hbar\gamma = 4$, Fig.~\ref{fig-sb-rates}A--F, a less quantum, overdamped case, $\beta\hbar\Omega = 1$, $\beta\hbar\gamma = 16$, Fig.~\ref{fig-sb-rates}G--I and also for a classical Debye spectral density $\beta\hbar\omega_\mathrm{D} = 0.5$, Fig.~\ref{fig-sb-rates}J--L. 

\highlight{In the top row of Fig.~\ref{fig-sb-rates} (A--C) the ratio of $\epsilon$ to $\lambda$ is held fixed at 2 and $\lambda$ is systematically increased for an underdamped Brownian Oscillator spectral density, and we see that the the OGR method remains accurate across the range of reorganisation energies. This accuracy is retained as ratio of $\epsilon$ to $\lambda$ is varied at fixed $\lambda$ for this spectral density, going from the Marcus activation-less case $\epsilon = \lambda$, to deep in the inverted regime $\epsilon = 5 \lambda$, as can be seen in the second row of Fig.~\ref{fig-sb-rates} (D--F). The accuracy of the OGR method also holds in the overdamped case and for a Debye spectral density as $\epsilon/\lambda$ is varied, as is shown in the third and fourth rows of Fig.~\ref{fig-sb-rates} (G--I and J--L) respectively, although the OGR is less accurate at lower values of $\epsilon/\lambda$ for the overdamped Brownian oscillator.  

In all cases we see that for large values of $\beta \Delta$ the GR rate can overestimate the exact rate by up to four orders of magnitude, whereas the OGR has a maximum error of a factor of $\sim 4$, which occurs for the overdamped Brownian oscillator spectral density at intermediate values of the coupling and low values of $\epsilon/\lambda$.} 
The turnover behaviour of the rate constant occurs at larger $\beta \Delta$ deeper in the inverted regime, and the OGR method also appears more accurate in this regime. The small errors at intermediate values of $\beta \Delta$ may have their origin in recrossing effects which are neglected in the OGR method. Despite this the overall near-quantitative accuracy of the OGR method for these spin-boson models is very encouraging, and OGR rate is never less accurate than the uncorrected GR rate. \highlight{A comparison of these results with the modified Marcus theory Eq.~\eqref{eq-mod-marcus} and Zusman theory\cite{zusman_theory_1988} are shown in Appendix \ref{app-zus}, where we see that Zusman theory fails for these spin-boson models and modified Marccus theory (which neglects nuclear quantum effects) can only qualitively predict the turnover behaviour.}
\begin{figure}
    \centering
    \includegraphics[width=0.44\textwidth]{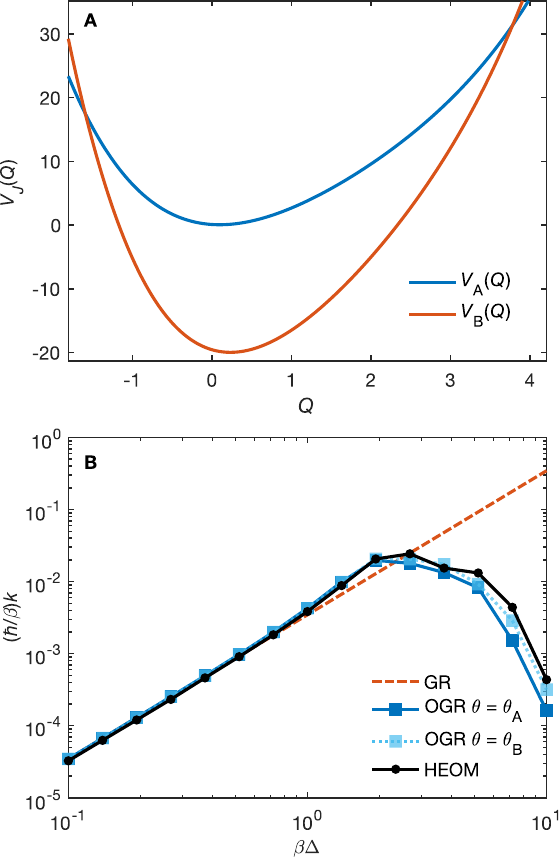}
    \vspace{-10pt}
    \caption{\footnotesize A) The diabatic potential curves for the A and B states used in the quartic anharmonic model. B) Rate constants calculated from HEOM for the quartic anharmonic model using GR and OGR methods with $\theta = \theta_\A$ (the proposed optimal solution) and $\theta = \theta_\B$, as well as the full HEOM results.}
    \label{fig-anharm-rates}
    \vspace{-15pt}
\end{figure}

\vspace{-20pt}
\subsection{Anharmonic model}
\vspace{-10pt}
In order to test the robustness of the OGR method to asymmetric frequencies and anharmonicity in the diabatic potential energy surfaces, exact and OGR rates were calculated for a model with quartic reaction coordinate, coupled to a bath to mimic a condensed phase environment. The model potentials are given by
\begin{align}
\begin{split}
    V_J(Q,\vb*{q}) &= \frac{1}{2}M\omega_J^2\bigg((Q-Q_{0,J})^2 - \alpha_J (Q-Q_{0,J})^3 \\
    &+ \frac{7}{12}(Q-Q_{0,J})^4\bigg) + V_{\mathrm{sb}}(Q,\vb*{q}) + \epsilon_J
\end{split}\\
V_{\mathrm{sb}}(Q,\vb*{q}) &= \sum_{\alpha} \frac{1}{2}m_\alpha \omega_\alpha^2 \left({q}_\alpha - Q \frac{c_{\alpha}}{m_\alpha \omega_\alpha^2}\right)^2,
\end{align}
Full quantum mechanical rate constants and the OGR rates (using the simplified version Eq.~\eqref{eq-kOGR2}) can be evaluated for this model with HEOM. Working in units where $\beta = \hbar = M = 1$, the model parameters are set to $\omega_\A = 2.5$,  $\omega_\B = 4$, $\alpha_\A = \alpha_\B = 0.25$, $Q_{0,\A} = 0$, $Q_{0,\B} = 0.25$, $\epsilon_\A = 0$, $\epsilon_\B = -20$. The harmonic bath is set as a Debye bath with $\lambda = 0.5$ and $\omega_\mathrm{D} = 2$. The diabatic potentials are shown in Fig.~\ref{fig-anharm-rates}A. Further details of the calculations are given in Appendix~\ref{app-heom}.
\begin{figure}[t!]
    \centering
    \includegraphics[width=0.46\textwidth]{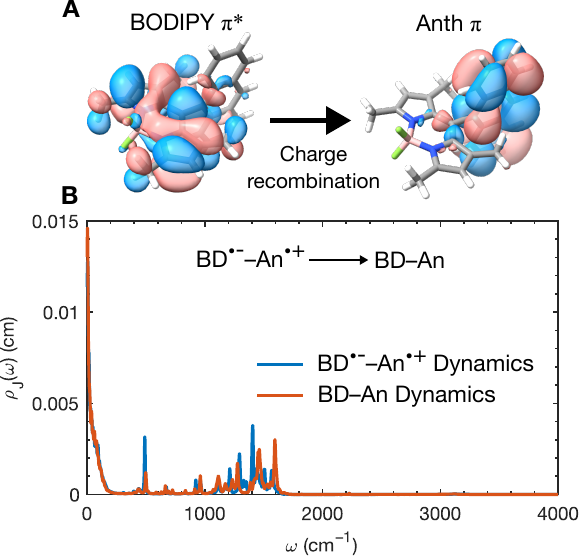}
    \vspace{-10pt}
    \caption{\footnotesize A) The BODIPY (boron-dipyrromethane)-Anthracene molecule, showing the orbitals involved in charge recombination from the excited \ce{BD^{$\bullet -$}-An^{$\bullet +$}} state to the \ce{BD-An} ground-state. B) Spectral densities calculated for this transition using molecular dynamics on both the \ce{BD^{$\bullet -$}-An^{$\bullet +$}} and \ce{BD-An} surfaces. Further details of the spectral density calculations are given in Ref.~\onlinecite{fay_unraveling_2024}.}
    \label{fig-bdan}
    \vspace{-15pt}
\end{figure}

\highlight{The OGR rate constant, which is optimised at $\theta = \theta_\A$, together with the $\theta = \theta_\B$ OGR rate, the GR prediction, and the full HEOM rates are shown in Fig.~\ref{fig-anharm-rates}B as a function of $\beta\Delta$.} In this example the exact rate constant agrees with the GR rate up to about $\beta\Delta = 0.7$, which is followed by a small increase in the rate with respect to the GR result up to $\beta\Delta = 2$. Beyond this value of $\beta\Delta$ the rate decreases with increasing $\Delta$, dropping to about $10^{-3} k_\mathrm{GR}$ at the largest $\Delta$ considered, $\beta\Delta = 10$. The OGR method still yields accurate rate constants in the weak-coupling limit $\beta \Delta < 1$, and it correctly captures the small increase in the rate constant up to $\beta\Delta = 2$ followed by the decrease in rate for larger $\Delta$. The largest error in this case is a factor of $\sim\!3$ error at $\beta\Delta \approx 7$, where the GR rate overestimates the rate constant by a factor of 40. The fact that the larger value of the OGR rate at $\theta = \theta_\B$ appears to be more accurate in this example at larger values of $\beta\Delta$ suggests that the simple minimization of the rate in the range $\theta_\A\leq \theta\leq \theta_\B$ may not always be the best method for optimising $\theta$, although it is still a considerable improvement over the GR rate. Alternative optimization strategies will be explored in future work.

\vspace{-20pt}
\section{Application: Strong-coupling effects in BODIPY-Anthracene charge recombination}
\vspace{-10pt}
\begin{table}[b]
    \centering
    \begin{tabular}{cccc}
        \hline
         & $\ce{BD^{\bullet -}-An^{\bullet +}}$ MD & $\ce{BD-An}$ MD & Average\\
         \hline
        GR & \qty{2.9d7}{s^{-1}} & \qty{4.4d7}{s^{-1}} & \qty{3.6d7}{s^{-1}} \\
        OGR & \qty{0.74d7}{s^{-1}}& \qty{1.2d7}{s^{-1}} & \qty{0.95d7}{s^{-1}}\\
        \hline
    \end{tabular}
    \caption{\footnotesize GR and OGR rate constants for the BODIPY-Anth charge recombination calculated with the different spectral densities and their average.}
    \label{tab-bdan-rates}
\end{table}

Having established that OGR theory provides a good description of non-adiabatic rates, particularly for spin-boson models deep in the Marcus inverted regime, an application of the OGR theory to charge recombination in a BODIPY-anthracene triplet photosensitizer is considered. This molecule forms a BODIPY (BD) to anthracene (An) charge transfer state after photo-excitation, which can either undergo intersystem crossing to the triplet manifold or recombine to the ground electronic state.\cite{buck_spin-allowed_2019,wang_insights_2019,fay_unraveling_2024} Previous work has found that the $\ce{BD^{\bullet -}-An^{\bullet +} -> BD-An }$ charge recombination in acetonitrile process lies deep in the inverted regime, with $-\Delta A \approx 5 \lambda$,\cite{buck_spin-allowed_2019,wang_insights_2019,fay_unraveling_2024} and furthermore the good overlap between the BD $\uppi^*$ and An $\uppi$ orbitals means the electronic coupling is very large at $\qty{1904}{cm^{-1}}\approx 9.2\ k_{\mathrm{B}}T$ (at $T=\qty{298}{K}$).\cite{fay_unraveling_2024} The orbitals involved and the \ce{BD-An} molecule are shown in Fig.~\ref{fig-bdan}A. 

In a previous study\cite{fay_unraveling_2024} a spin-boson mapping was constructed for the charge recombination process based on molecular dynamics simulations using molecular mechanics force-fields parameterised for the ground and charge-transfer states. The Fourier transform of the energy gap correlation function yielded a spectral density, which can be computed either for dynamics on the \ce{BD-An} or $\ce{BD^{\bullet -}-An^{\bullet +}}$ potential energy surface. These spectral distributions, $\rho(\omega) = \mathcal{J}(\omega) / (\omega\pi \lambda)$, are given in Fig.~\ref{fig-bdan}. Further details of these calculations used to parameterise the spin-boson models are given in Ref.~\onlinecite{fay_unraveling_2024}. Using these spectral densities, the rate constant for the charge recombination process can be estimated using either the standard GR approach or the OGR theory. \highlight{A comparison with a previously constructed model spectral density\cite{fay_unraveling_2024} given in Appendix \ref{app-bdan-model} suggests that this problem should be in a regime where OGR theory is reliable.} The GR and OGR rates \highlight{(with full optimisation, Eq.~\eqref{eq-kOGR1})} for both spectral densities are given in Table~\ref{tab-bdan-rates}, where one can see that for both spectral densities the rate drops by a factor of 4 on including strong diabatic coupling effects. 

The charge recombination rate plays an important role in determining the triplet quantum yield of \ce{BD-An}, because charge recombination competes with intersystem crossing from the singlet $\ce{BD^{\bullet -}-An^{\bullet +}}$ to the triplet excited state. By taking the charge recombination rate as the average of the OGR rates for the two spectral densities, the triplet quantum yield $\Phi_\mathrm{T}$ for BODIPY-Anthracene can be calculated. Without strong coupling effects in the charge recombination, $\Phi_\mathrm{T}$ was previously calculated to be 0.8, however including strong coupling effects increases this to 0.88, much closer to the experimental value of 0.93\cite{buck_spin-allowed_2019}--0.96.\cite{wang_insights_2019} (Details of this calcualtion are given in Appendix~\ref{app-phiT}) This highlights the importance of strong electronic coupling effects in a real molecular system. Further details of the other photophysical processes and the triplet yield calculation are given in Ref.~\onlinecite{fay_unraveling_2024}.
\vspace{-25pt}
\section{Conclusions}
\vspace{-10pt}
In this paper I have presented an Optimal Golden Rule theory for calculating rate constants of non-adiabatic processes in the Marcus inverted regime where coupling between diabatic electronic states becomes large. The theory is based on rotating to a new set of diabatic states where the strongly coupled states are partially mixed, in such a way that the coupling between states is reduced, enabling pertubation theory to be applied more accurately. This simple approximation works well for a range of tested spin-boson models and an anharmonic model. OGR theory can be applied straightforwardly with existing methods for computing GR rates for anharmonic potential energy surfaces, such as using spin-boson mappings, instanton approaches and path integral methods. The simplicity of OGR theory also allows one to obtain a generalization of Marcus theory and MLJ theory to strong electronic couplings in the inverted regime, where it was found the OGR theory predicts that strong electronic coupling effectively renormalises the diabatic coupling, free energy change and reorganisation energy of the reaction. 

Evaluation of the OGR rate is equivalent to the evaluation of a GR rate, so any method that can be applied with GR rates can also in principle be combined with OGR theory. This will enable OGR theory to be applied to atomistic models of reactions, beyond spin-boson models and model potentials considered in this work. The current simple condition for optimising the mixing angle $\theta$ just requires evaluation of expectation values of quantities in the original A and B diabatic equilibrium states, which can also be done very straightforwardly for atomistic models including nuclear quantum effects using path integral methods.\cite{markland_nuclear_2018} \highlight{Despite its simplicity, there may exist better $\theta$ optimisation strategies for obtaining accurate rate constants. For example it is possible that a more direct minimisation of $\ev{\Delta_{+-}^2}_\pm$ or a minimisation of the initial value of the flux-flux correlation function ${\Tr}[\hat{F}^2 e^{-\beta \hat{H}_\pm}] \propto Q_{\pm} \ev{\Delta_{+-}^2}_\pm$ (as is done in the NAQI method\cite{lawrence_general_2020} or transition state theory) may provide more accurate results for the OGR method. Furthermore there is no reason that the same value of $\theta$ should be used to evaluate both the approximate flux-flux correlation function and the reactant partition function, and these could be optimised separately, although early explorations of this idea have not produced better results than those presented here. Thorough investigation of these alternative approaches will be the subject of future work.} 
Although there may still be room for improvement, OGR theory already enables the accurate calculation of non-adiabatic rate constants at arbitrary diabatic coupling strengths in the inverted regime, a regime currently inaccessible to other methods such as the interpolation formula approach,\cite{lawrence_calculation_2019} and it shows considerable promise in application to atomistic models and real-world molecular systems. OGR theory also offers a route to generalising theory of other non-adiabatic processes, such as proton-coupled electron transfer,\cite{hammes-schiffer_theory_2010} excitation energy transfer,\cite{renger_theory_2009} both of which are relevant in enzymatic and photosynthetic systems, as well as electrochemical processes.\cite{marcus_theory_1965,santos_models_2022}

\vspace{-10pt}
\subsection*{Acknowledgements}
\vspace{-10pt}
\noindent I would like to thank Joseph Lawrence for useful discussions as well as David Limmer and the Limmer group for some early conversations on this work. This work was supported by ``Agence Nationale de la Recherche'' through the project MAPPLE (ANR-22-CE29-0014-01). 
\vspace{-10pt}
\subsection*{Data availability}
\vspace{-10pt}
\noindent All data is provided within the article itself. Code used to perform the HEOM calculations is available at \url{https://github.com/tomfay/heom-lab} and scripts are available from the corresponding author upon request.
\vspace{-10pt}
\subsection*{Conflicts of interest}
\vspace{-10pt}
\noindent The author declares no conflicts of interest.

\appendix

\vspace{-10pt}
\section{Derivation of simplified $k_{\mathrm{GR},\theta}$}\label{app-kOGR-simp}
\vspace{-10pt}
The coupling between $\ket{+}$ and $\ket{-}$ states involves a term proportional to $\hat{U} = \hat{H}_+ - \hat{H}_-$. In the static approximation or classical limit, where the kinetic energy and potential energy operators commute, the time integral of the propagators becomes,\cite{sparpaglione_dielectric_1988}
\begin{align}
    \int_{-\infty}^{+\infty} \dd{t} e^{+i\hat{H}_-t/\hbar} e^{-i\hat{H}_-t/\hbar} \approx 2\pi \hbar \delta(\hat{U}),
\end{align}
and therefore it is clear that these terms in the rate constant proportional to $\hat{U}^n$ vanish. We can also show this more generally. First  considering the terms where a single $\hat{U}$ operator appears, we note that the integrand can be written as a derivative as follows
\begin{align}
    \int_{-\infty}^{\infty} &\Tr_\mathrm{n}[e^{-\beta \hat{H}_+ -i \hat{H}_+ t/\hbar}\hat{\Delta}e^{+i \hat{H}_- t/\hbar}\hat{U}]\dd{t} \nonumber \\
    &=i\hbar\int_{-\infty}^{\infty} \dv{t}\Tr_\mathrm{n}[e^{-\beta \hat{H}_+ -i \hat{H}_+ t/\hbar}\hat{\Delta}e^{+i \hat{H}_- t/\hbar}]\dd{t}\\
    &= i\hbar\left[\Tr_\mathrm{n}[e^{-\beta \hat{H}_+ -i \hat{H}_+ t/\hbar}\hat{\Delta}e^{+i \hat{H}_- t/\hbar}] \right]_{t=-\infty}^{t= \infty}
\end{align}
and since this correlation function decays to zero, these contributions are zero. The same argument applies to the terms where $\hat{\Delta}$ and $\hat{U}$ are swapped. Applying the same argument twice we find that the terms involving two factors of $\hat{U}$ are also zero, because
\begin{align}
    &\Tr_\mathrm{n}[e^{-\beta \hat{H}_+ -i \hat{H}_+ t/\hbar}\hat{U}e^{+i \hat{H}_- t/\hbar}\hat{U}] \nonumber \\
    &= (i\hbar^2) \dv[2]{t} \Tr_\mathrm{n}[e^{-\beta \hat{H}_+ -i \hat{H}_+ t/\hbar}e^{+i \hat{H}_- t/\hbar}].
\end{align}
Overall this shows that Eq.~\eqref{eq-ogr} simplifies to Eq.~\eqref{eq-ogr-simp}.


\vspace{-10pt}
\section{HEOM calculations}\label{app-heom}
\vspace{-10pt}
Numerically exact rate constants for the models considered are obtained using the hierarchical equations of motion (HEOM).\cite{tanimura_time_1989,tanimura_numerically_2020} The \texttt{heom-lab} code was used to perform all simulations, using a Matsurbara expansion for constructing the hierarchy with the low temperature correction from Ref.~\onlinecite{fay_simple_2022}. The HEOM truncation scheme from Ref.~\onlinecite{lindoy_quantum_2023} used in all calculations. Up to $M=3$ Matsubara modes were included explicitly in the calculations with the depth of the hierarchy set to between $L=40$ and $L=150$ depending on the spectral density used.

The rate constant was extracted by initialising the system in the $\hat{\rho}(0) = \dyad{\A}e^{-\beta\hat{H}_{\A}}/Q_{\A}$ state and fitting the population dynamics of state B to $p_\B(t) = (p_{\B,\infty}-p_{\B,0})(1-e^{-k_\mathrm{obs}t}) + p_{\B,0}$ after initial transient oscillations had decayed. Strictly speaking $k_\mathrm{obs}$ is the sum of forward and backward rates, but in all examples considered the ratio of the backward to forward rates is $\lesssim 10^{-9}$ so the rate constant is taken to be the fitted value $k_\mathrm{obs}$.

For the anharmonic model the system Hamiltonian was constructed in a Colbert-Miller DVR basis\cite{colbert_novel_1992} with 500 basis functions for each diabatic state and then diagonalised. The $N_\A=11$ lowest energy states for diabat A and $N_\B=15$ lowest energy states for diabat B were used to construct the full system Hamiltonian for the exact rate calculations and the $\Delta = 0$ system Hamiltonian for the GR calculations with HEOM. 
\begin{SCfigure*}[0.72]
\vspace{-10pt}
    \includegraphics[width=0.72\textwidth]{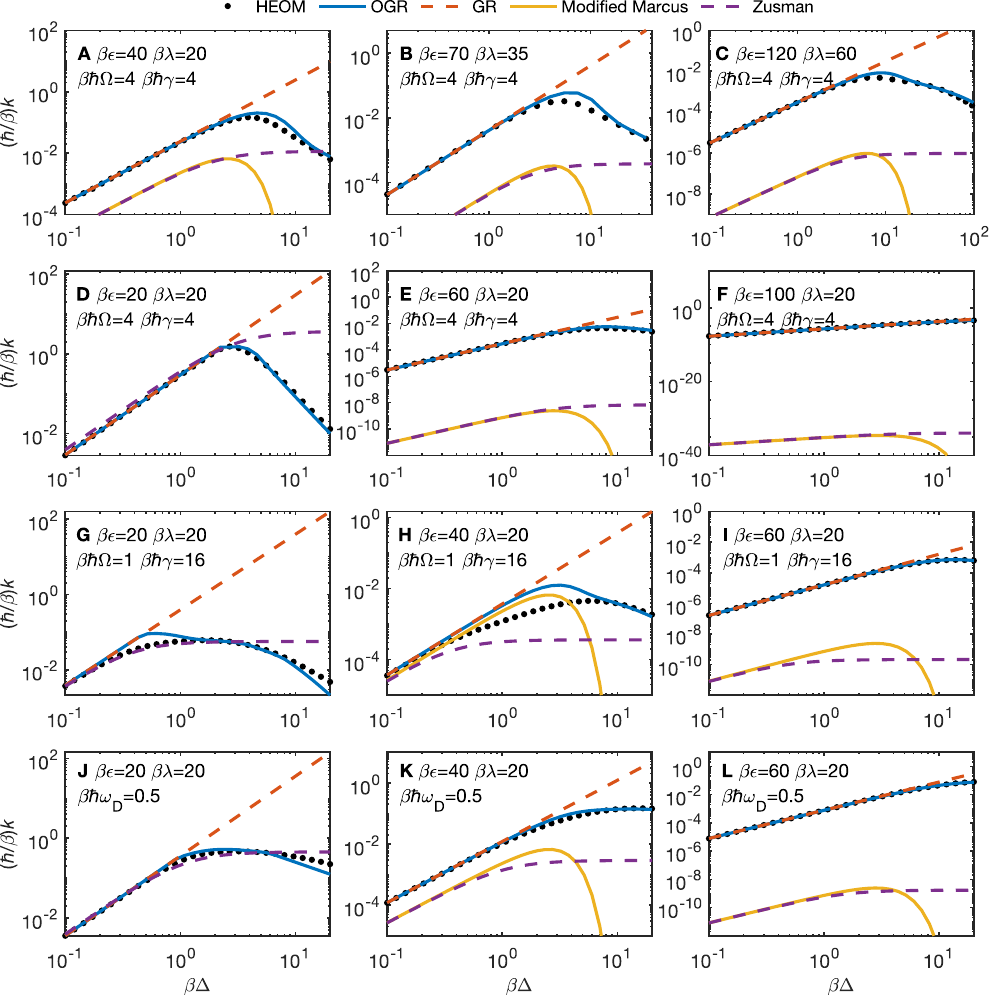}
    \vspace{-10pt}
    \caption{\footnotesize A comparison of HEOM spin-boson model rates together with OGR and GR rates and the modified Marcus and Zusman theory rates. All model parameters are the same as in Fig.~\ref{fig-sb-rates}.}\label{fig-zus}
    
\end{SCfigure*}
\highlight{
\vspace{-10pt}
\section{Comparison of Modified Marcus and Zusman theories}\label{app-zus}
\vspace{-10pt}

Here a comparison of modified Marcus theory and Zusman theory are shown for the spin-boson models considered in Sec.~\ref{sec-sb-models}. Zusman theory gives the rate constant as\cite{zusman_theory_1988,sparpaglione_dielectric_1988,sparpaglione_dielectric_1988-1}
\begin{align}
    k_\mathrm{Zus} = \frac{k_{\mathrm{MT,f}}}{1 + \tau_{\mathrm{S,f}}k_{\mathrm{MT,f}} + \tau_{\mathrm{S,b}}k_{\mathrm{MT,b}} }
\end{align}
where $k_{\mathrm{MT,f/b}}$ are the forward/backward Marcus theory rate constants, and $\tau_{\mathrm{S,f/b}}$ are effective solvent relaxation times given by,
\begin{align}
\begin{split}
    &\tau_{\mathrm{S,f/b}} = \tau_\mathrm{S}\bigg(\ln(2) \\
    &+ \frac{(\lambda \pm \Delta A)^2}{2k_\mathrm{B} T \lambda } {}_2F_2\left(1,1;\frac{2}{2},2;\frac{(\lambda \pm \Delta A)^2}{4k_\mathrm{B} T \lambda }\right) \bigg)
\end{split}
\end{align}
where ${}_2F_2(a_1,a_2;b_1,b_2;z)$ is a generalised hypergeometric function, and $\tau_\mathrm{S}$ is the relaxation time of the energy gap coordinate. Qualitatively Zusman theory predicts that the rate constant plateaus at large values of $\Delta$, which is clearly not what is observed in the spin-boson models in Sec.~\ref{sec-sb-models} which show turnover behaviour. For the Debye spectral density $\tau_\mathrm{S} = 1/\omega_\mathrm{D}$, for the underdamped Brownian Oscillator $\tau_\mathrm{S} = 2 / \gamma $ and for the overdamped Brownian Oscillator $\tau_\mathrm{S} = \gamma/ \Omega^2$.

In Fig.~\ref{fig-zus} we show the spin-boson model results from Fig.~\ref{fig-sb-rates} together with the Modified Marcus and Zusman results (note that the modified Marcus rate is not defined for the activation-less cases: Fig.~\ref{fig-sb-rates}D, G and J). We see that neither is quantitatively accurate; modified Marcus theory generally over-predicts the drop in the rate constant at large values of $\Delta$ whilst Zusman theory simply plateaus at large $\Delta$. Furthermore, since both are classical theories the neglect of nuclear quantum effects leads to large errors even in the Golden Rule regime. We see that modified Marcus theory at least roughly predicts the value of $\Delta$ at which rate turnover occurs relatively accurately, so it should at least be useful as a guide for when GR theory is likely to be accurate. The importance of nuclear quantum effects in the inverted regime with large diabatic couplings clearly means that accounting for the details of the full spectral density are essential for quantitatively accurate rate calculations in this regime.
}

\highlight{
\vspace{-10pt}
\section{OGR with the BD-An model spectral density}\label{app-bdan-model}
\vspace{-10pt}

In Ref.~\onlinecite{fay_unraveling_2024} an approximate post-GR correction to the BD-An charge recombination rate was calculated using a model spectral density, which is a sum of a Debye spectral density (with $\beta\lambda = 10.1459$ and $\beta\hbar\omega_\mathrm{D} = 0.1831$) and an underdamped Brownian Oscillator spectral density (with $\beta\lambda = 8.6780$, $\beta\hbar\Omega = 6.76$ and $\beta\hbar\gamma = 4$). The model spectral distribution is shown in Fig.~\ref{fig-bdan-model}A. This model predicted a much smaller reduction in the rate, by only about a factor of 0.9, relative to the GR rate, comapred to the OGR calculations with the full spectral density performed here. However this simplified model overestimates the GR rate compared to the full spectral density by over a factor of 1000. This is because of the polynomial decaying tails of these spectral densities which leads to very large tunnelling effects. Despite this issue with the model, we can use the HEOM results to benchmark OGR theory. This comparison is shown in Fig.~\ref{fig-bdan-model}B, where we see that OGR theory is very accurate for this model (note that for BD-An $\beta\Delta \approx 10$). This suggests that OGR theory should be reliable for the full spetral density as well.
\begin{figure}[b]
    \centering
\vspace{-10pt}
    \includegraphics[width=0.42\textwidth]{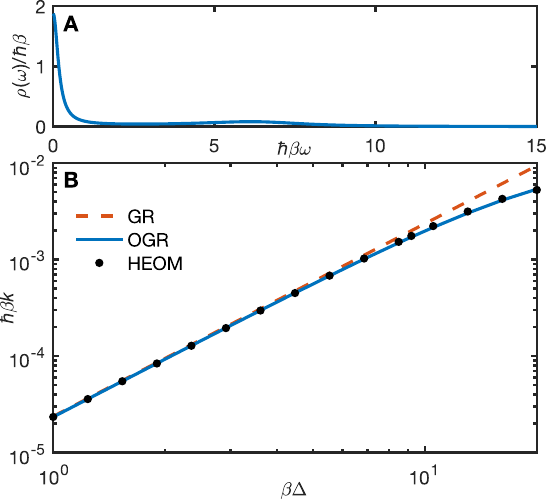}
    \vspace{-10pt}
    \caption{\footnotesize A) The model spectral distribution for charge recombination in BD-An. B) A comparison of GR, HEOM and OGR rates with this model as a function of $\Delta$.}
    \label{fig-bdan-model}
    \vspace{-10pt}
\end{figure}
}
\vspace{-10pt}
\section{Triplet yield calculation for BODIPY-Anthracene}\label{app-phiT}
\vspace{-10pt}
The BODIPY-Anthracene singlet charge transfer state can either recombine to the ground singlet electronic state, or undergo intersystem crossing to yield the excited triplet state, which can be used as a photosensitizer. This charge transfer state is in pre-equilibrium with the \ce{BD^*-An} valence excited state, so the overall triplet yield $\Phi_\mathrm{T}$ is
\begin{align}
    \Phi_\mathrm{T} = \frac{p_\mathrm{CT}  k_{\mathrm{ISC,tot}}}{p_\mathrm{CT} k_{\mathrm{ISC,tot}} + p_\mathrm{CT} k_{\mathrm{CR}} + k_{\mathrm{Fl,tot}}}. 
\end{align}
${k_{\mathrm{ISC,tot}}}$ is the sum of intersystem crossing rates from the charge transfer (CT) state, $k_{\mathrm{CR}}$ is the charge recombination rate back to the ground state (which is recalculated in this work), $k_{\mathrm{Fl,tot}}$ is the total effective fluorescence rate from both the CT and \ce{BD^*-An} states, and $p_\mathrm{CT} = e^{-\Delta A_\mathrm{CS}/k_\mathrm{B}T} / (1 + e^{-\Delta A_\mathrm{CS}/k_\mathrm{B}T}) $ is the pre-equilibrium population of the CT state ($\Delta A_\mathrm{CS}$ is the free energy change of excited state charge separation). From Ref.~\onlinecite{fay_unraveling_2024} $k_\mathrm{ISC,tot} = \qty{2.05e8}{s^{-1}}$, $k_\mathrm{Fl,tot} = \qty{1.74e7}{s^{-1}}$ and $p_{\mathrm{CT}} = 0.903$. 
\vspace{-10pt}
\section*{References}
\vspace{-15pt}
\bibliography{main-refs}

\end{document}